\documentclass[fonts]{icst}

\usepackage{moreverb}

\usepackage[breaklinks,colorlinks,bookmarksopen,bookmarksnumbered,linkcolor=ICSTblue,citecolor=blue,urlcolor=ICSTblue]{hyperref}
\usepackage{breakurl}

\usepackage{amssymb}
\usepackage{graphicx}
\usepackage[ruled,vlined,linesnumbered]{algorithm2e}
\usepackage{float}
\usepackage{color}
\usepackage{soul}
\usepackage{amsmath}
\usepackage{caption}
\usepackage{makecell}

\usepackage{url}

\newcommand\BibTeX{{\rmfamily B\kern-.05em \textsc{i\kern-.025em b}\kern-.08em
T\kern-.1667em\lower.7ex\hbox{E}\kern-.125emX}}

\def\volumeyear{2014}

\journalname{XXXXXX}
\articletype{Research Article}
 \def\volumeyear{XXXX}
 \def\volumenumber{XX}
  \def\issuemonth{XX-XX}
  \def\issuenumber{X}
  \def\articleid{XX}
  \copyrightnote{This is an open access article distributed under the terms of the Creative Commons Attribution license (\url{http://creativecommons.org/licenses/by/3.0/}), which permits unlimited use, distribution and reproduction in any medium so long as the original work is properly cited.}
  \def\articledoi{10.4108/XX.X.X.XX}
\received{XXXX}
  \accepted{XXXX}
  \published{XXXX}

\begin{document}

\runningheads{Iqbal H. Sarker}{Understanding the Role of Data-Centric Social Context in Personalized Mobile Applications}

\title{Understanding the Role of Data-Centric Social Context in Personalized Mobile Applications}

\author{Iqbal H. Sarker \affil{*}}

\address{Department of Computer Science and Software Engineering, \\ Swinburne University of Technology, \\ Melbourne, VIC-3122, Australia.}

\abstract{Context-awareness in personalized mobile applications is a growing area of study. Social context is one of the most important sources of information in human-activity based applications. In this paper, we mainly focus on \textit{social relational context} that represents the \textit{interpersonal relationship} between individuals, and the \textit{role} or influence of such context on users' diverse phone call activities in their real world life. Individuals different phone call activities such as making a phone call to a particular person or responding an incoming call may differ from person-to-person based on their interpersonal relationships such as family, friend, or colleague. However, it is very difficult to make the device understandable about such semantic relationships between individuals and the relevant context-aware applications. To address this issue, in this paper, we explore the \textit{data-centric} social relational context that can play a significant role in building context-aware personalized mobile applications for various purposes in our real world life.}

\keywords{Mobile phones, phone log data, user activity, personalization, contexts, social context, interpersonal relationship, mobile applications, call interruptions, call reminders, context-aware systems.}

\tnotetext[1]{This article is a preprint version of the journal \textit{"EAI Endorsed Transactions on Context-aware Systems and Applications"}. \\}

\fnotetext[1]{Corresponding author: Iqbal H. Sarker.  Email: \email{msarker@swin.edu.au}}

\maketitle

\section{Introduction}
\label{Introduction}
Nowadays, mobile phones have become part of our life. Mobile phones have become one of the primary ways, in which people around the globe communicate with each other for various purposes. According to ITU (International Telecommunication Union), cellular network coverage has reached 96.8\% of the world population, and this number even reaches 100\% of the population in the developed countries \cite{number-of-mobile-phone-users}. Mobile phones have transformed from merely communication tools to smart and highly personal important devices of individual users, which are able to assist them in a variety of their daily activities in day-to-day situations. These smart mobile phones have incorporated a variety of significant and interesting features to facilitate various context-aware personalized services, such as context-aware interruption management or reminder system, for the benefit of end mobile phone users in their daily life in a context-aware computing environment. 

In recent times, the smart phones are becoming more and more powerful in both computing and the data storage capacity. For the purpose of mobile communication, people typically make a phone call to a particular person or respond an incoming call from a person. The mobile phones have the ability to log such phone call activities of individual mobile phone users in various contexts, such as temporal \cite{sarker2016phone}, spatial or user location \cite{eagle2006reality}. In addition to this spatio-temporal context, the \textit{social relational context} or \textit{interpersonal relationship} between individuals has also an influence on their phone call activities \cite{sarker2016understanding}. The reason is that individuals phone call activities may differ from person-to-person based on their interpersonal relationships, such as family, friend, or colleague etc.

Let's consider an example for a particular mobile phone user, Alice. She works in a corporate office as an executive officer. In the morning, she attends a regular meeting in her office on Monday. Typically, she rejects the incoming phone calls during that time period as she does not want to be interrupted with phone calls during the meeting. The reason is that the interruptions may not only create disturbance for herself but also for the surrounding other people in the meeting. However, if the phone call comes from her boss, she wants to answer the call as it likely to be important for her, even though she is in a meeting. Hence, the social relational context $`interpersonal \; relationship \rightarrow boss'$ has an influence to make her phone call decision. Similarly, her phone call decisions may also vary for some other relationships such as friend, family or colleague, in her real world life.

In the real world, people live in different societies and maintain their interpersonal relationships according to their own cultures. For instance, in a family relationship, one may call her mother `mom', while another may call as `mammy'. Thus it is very difficult to make the device understandable about such semantic relationships between individuals and the relevant context-aware applications. In contrast, mobile phone data recorded in the device logs can be a rich resource to analyze the data-centric context, in which we are interested in. For instance, phone call log data including individuals' unique phone numbers offers the potential to understand person-to-person phone call activities and corresponding relationships for those activities. Based on this, we aim to explore the \textit{data-centric} social relational context and the role of this context in building various personalized context-aware applications.

The main contribution of this paper is - we explore the potentiality of \textit{data-centric social context}, i.e., how individuals mobile phone data can be used as a social context in personalized mobile applications. For this purpose, we first discuss various categories of relationships in our societies and their impacts and issues on individuals phone call activities. We also discuss how the \textit{data-centric} social relational context or the \textit{interpersonal relationship} between individuals can play a significant role in building context-aware personalized mobile applications.

\begin{table*}
	\centering
	\caption{Examples of mobile phone data.}
	\label{sample-time-seris-datasets}
	\begin{tabular}{|c|c|c|c|c|} 
		\hline
		\bf \makecell{Phone Call Information \\ (Call Type, Phone Number with Relationship, Call Duration)} & \bf \makecell{Phone Call \\ Activity}\\  
		\hline
		Incoming call from 047XXXX231 (Mother), 565 Sec & Accept \\
		Incoming call from 047XXXX232 (Friend), 0 Sec & Reject \\  
		Incoming call from 047XXXX233 (Unknown), 0 Sec & Reject \\
		Missed call from 047XXXX234 (Friend), 0 Sec & Missed \\
		Make a phone call to 047XXXX231 (Mother), 500 Sec & Outgoing \\ 
		Incoming call from 047XXXX235 (Colleague), 120 Sec & Accept \\
		Incoming call from 047XXXX231 (Mother), 565 Sec & Accept \\
		Incoming call from 047XXXX232 (Friend), 0 Sec & Reject \\  
		Incoming call from 047XXXX233 (Unknown), 0 Sec & Reject \\
		Missed call from 047XXXX234 (Friend), 0 Sec & Missed \\ 
		Make a phone call to 047XXXX234 (Friend), 200 Sec & Outgoing \\  
		Incoming call from 047XXXX235 (Colleague), 120 Sec & Accept \\ 
		Incoming call from 047XXXX231 (Mother), 565 Sec & Accept \\
		Incoming call from 047XXXX232 (Friend), 0 Sec & Reject \\  
		Incoming call from 047XXXX233 (Unknown), 0 Sec & Reject \\
		Missed call from 047XXXX234 (Friend), 0 Sec & Missed \\  
		Incoming call from 047XXXX235 (Colleague), 120 Sec & Accept \\     
		\hline
	\end{tabular}
\end{table*}

\section{Call Activities and Social Relational Context}
\label{Interpersonal Relationship and Phone Call Activities}
In the real world, the common phone call activities of an individual mobile phone user are - (i) answering the incoming phone call by the user for a particular time period or duration, i.e., `Accept', (ii) instantly decline the incoming phone call by the user, i.e., `Reject', (iii) the phone rings for an incoming call but the user misses the call, i.e., Missed, and (iv) making a phone call to a particular person, i.e., `Outgoing'. However, all these phone call activities of an individual user are not similar to all, may vary from person-to-person, in their real world life. The reason is that the social relational context or interpersonal relationship between individuals has a strong influence in their phone call activities \cite{sarker2016understanding}. For instance, in a particular context, one may answer an incoming call for a specific relationship and may decline the call for a different relationship.

In the real world, the study of human relationships is of concern to three broad areas: (i) sociology, i.e., the study of the development, structure, and functioning of human society, (ii) psychology, i.e., the study of the mind and behavior of human, and (iii) anthropology, i.e., the study of humans and human behavior and societies in the past and present. Typically, human beings are naturally social creatures. They build interpersonal relationships among themselves by nature. Based on this, interpersonal relationships can be treated as social associations, connections, or affiliations between two or more people. According to Wikipedia \cite{WiKiRelation}, ``An interpersonal relationship is a strong, deep, or close association or acquaintance between two or more people that may range in duration from brief to enduring''. This association may be based on inference, love, solidarity, regular business interactions, or some other type of social commitment. Therefore, in case of phone call activities, the interpersonal relationship is the social relational bonding between the caller and callee. Based on this, we summarize a number of categories of interpersonal relationships that might have the impacts on individuals phone call activities. These are:

\begin{itemize}
	\item \textit{Family Relationship}: This is one of the most important interpersonal relationship in our societies. Typically, we all belong to a particular family by born or nature. Family is defined as a domestic group of people either by legally bond or blood bond. Legally bond in a family is related to marriages, adoptions, and guardianships, including the rights, duties, and obligations of those legal contracts. On the other hand, blood bond includes both close and distant relatives such as siblings, parents, grandparents, aunts, uncles, nieces, nephews, and cousins. The phone call activities of an individual may not be similar for all these family members, may differ from person-to-person. For instance, one's phone call activities with her mother may not be similar with her cousin, even though both are in a same family.
	
	\item \textit{Friendship Relationship}: Friendship is an unconditional interpersonal relationship in our societies or in the world. This is a social relationship where there is no specific formalities. Individuals may create friendship relationship between them according to their own choice or interests, and can enjoy one another's presence. An individual may have many friends. However, the phone call activities of an individual may not be similar to all her friends, may differ from friend-to-friend. For instance, one's phone call activities with her close friend may not be similar with her another friend or friend of friend (fof).
	
	\item \textit{Love or Romantic Relationship}: If an interpersonal relationship between individuals is characterized by passion, intimacy, trust and respect, then we call it love or romantic relationship. In the real world, individuals in a romantic relationship are deeply attached to each other and share a special bonding in their life, like boyfriend, girlfriend or other significant person. Thus, the phone call activities may differ accordingly based on their relationships.

	\item \textit{Professional or Work Relationship}: All the above relationships are the personal relationships between individuals. However, building professional relationships in an organization is a crucial thing in our daily activities. A meaningful and strong relationship at workplace leads to better output, enhancing the personal productivity of the employees. Professional relationships can be built up if the individuals work together for the same or different organizations. An individual may have professional relationship with many people. However, the phone call activities of an individual may not be similar to all her colleagues or co-workers, may differ from colleague-to-colleague. For instance, one's phone call activities with her boss may not be similar with her another colleague. Thus the phone call activities may differ based on these particular relationships according to the importance or priorities in works.
	
	\item \textit{Unknown or Others}: Simply this category represents unknown people or others with whom there is no specific relationship. Though, the people in this category have no significant value in terms of interpersonal relationship, this category can be a major part in terms of population. Thus individuals in this category might have an impact on phone call activities in our real life. The reason is that by using a particular phone number one can easily communicate with another one in the world. For instance, one may receive a phone call from an unknown person or a person with whom there is no specific relationship. Thus the phone call activities with unknown people may not be similar with the people having a particular interpersonal relationship discussed above.
\end{itemize}

The above discussions highlight that individuals phone call activities are not identical to all, may vary from person-to-person depending on their social relational context or interpersonal relationship between them. Thus we motivated to one-to-one relationship between individuals in their phone call activities and relevant personalized context-aware applications to assist themselves effectively and intelligently. In order to support this line, in this work, we explore the \textit{data-centric} social relational context that represents the one-to-one unique relationship between individuals according to their mobile phone data.

\begin{table*}
	\begin{center}
		\caption{Sample examples of phone numbers in mobile phone data and corresponding social relational context}
		
		\label{sample-relationships}
		\begin{tabular}{l|c|c|c} 
			\textbf{\makecell{Phone Call \\ Activity}} & \textbf{\makecell{Individuals \\ Phone Number}} & \textbf{\makecell{Relationship \\ (data-centric)}} & \textbf{\makecell{Relationship \\ (semantic)}}\\
			\hline
			\makecell{Accept} & \makecell{047XXXX231} & \makecell{$Rel_{A}$} & \makecell{Mother}\\
			\hline
			\makecell{Reject} & \makecell{047XXXX232} & \makecell{$Rel_{B}$} & \makecell{Friend (close)}\\
			\hline
			\makecell{Reject} & \makecell{047XXXX233} & \makecell{$Rel_{C}$} & \makecell{Unknown}\\
			\hline
			\makecell{Missed} & \makecell{047XXXX234} & \makecell{$Rel_{D}$} & \makecell{Friend (best)}\\
			\hline
			\makecell{Outgoing} & \makecell{047XXXX231} & \makecell{$Rel_{A}$} & \makecell{Mother}\\
			\hline
			\makecell{Accept} & \makecell{047XXXX235} & \makecell{$Rel_{E}$} & \makecell{Colleague}\\
			\hline
			\makecell{Accept} & \makecell{047XXXX231} & \makecell{$Rel_{A}$} & \makecell{Mother}\\
			\hline
			\makecell{Reject} & \makecell{047XXXX232} & \makecell{$Rel_{B}$} & \makecell{Friend (close)}\\
			\hline
			\makecell{Reject} & \makecell{047XXXX233} & \makecell{$Rel_{C}$} & \makecell{Unknown}\\
			\hline
			\makecell{Missed} & \makecell{047XXXX234} & \makecell{$Rel_{D}$} & \makecell{Friend (best)}\\
			\hline
			\makecell{Outgoing} & \makecell{047XXXX234} & \makecell{$Rel_{D}$} & \makecell{Friend (best)}\\
			\hline
			\makecell{Accept} & \makecell{047XXXX235} & \makecell{$Rel_{E}$} & \makecell{Colleague}\\
			\hline
			\makecell{Accept} & \makecell{047XXXX231} & \makecell{$Rel_{A}$} & \makecell{Mother}\\
			\hline
			\makecell{Reject} & \makecell{047XXXX232} & \makecell{$Rel_{B}$} & \makecell{Friend (close)}\\
			\hline
			\makecell{Reject} & \makecell{047XXXX233} & \makecell{$Rel_{C}$} & \makecell{Unknown}\\
			\hline
			\makecell{Missed} & \makecell{047XXXX234} & \makecell{$Rel_{D}$} & \makecell{Friend (best)}\\
			\hline
			\makecell{Accept} & \makecell{047XXXX235} & \makecell{$Rel_{E}$} & \makecell{Colleague}\\
			\hline
		\end{tabular}
	\end{center}
\end{table*}

\section{Data-Centric Social Relational Context}
\label{Data-Centric Social Relationship Context}
In this section, we discuss how individuals' mobile phone data can be used as the data-centric social context. For this, we first discuss about the real life mobile phone datasets related to individuals phone call activities, and the relevant features exist in the datasets.

\subsection{Real-life Mobile Phone Data}
Recent advances in mobile phones and their powerful capabilities have enabled the collection of mobile phone data with various phone call activities of the users. For instance, `Reality Mining' is one of the most popular mobile phone dataset consists of 94 individual mobile phone users' phone call activities over nine months which were collected at Massachusetts Institute of Technology (MIT) by the Reality Mining Project [Massachusetts Institute of Technology 2007] \cite{eagle2006reality}. These 94 individuals are faculty, staff, and students. The datasets include people with different types of calling activities and corresponding call-related information. Another dataset is `Nodobo' \cite{bell2011nodobo} that includes call records of 27 participants over a 5-months period from September 2010 to February 2011. This dataset is gathered during a study of the mobile phone usage at University of Strathclyde. All the participants in their study were high-school students. Another dataset is `Swin' \cite{sarker2016behavior} that was collected directly by us using our data collecting application. Data was collected from 22 individual mobile users of different professions such as undergraduate students, post graduate students, university lecturers and industry professionals, from August 2014 to September 2015, at Swinburne University of Technology, Melbourne, Australia. All these datasets contain various phone call activities, e.g., INCOMING (Accept or Reject), MISSED and OUTGOING, and corresponding phone number of individuals, and others call related meta-data.

Table \ref{sample-time-seris-datasets} shows an example of mobile phone data of a sample user. It reports some pieces of information coming from such mobile phone data having various phone call activities with corresponding call related meta-data.

\subsection{Data-Centric Interpersonal Relationship}
As discussed in the earlier section, the semantic relationship categories like family, friend, love, professional or unknown, may not always be useful, for making phone call decisions in the real world in an automated mobile system. As individuals' phone call decisions may vary from person-to-person, we take into account \textit{unique identifier} of individuals for exploring \textit{data-centric} social relational context.

In the real world, a mobile phone number of an individual can be used to represent as the unique user identifier for mobile communication. For instance, one's phone number is well different with another one, even for a single digit in the number in order to avoid the conflicts in communication. Mobile phones automatically record the phone numbers in it's phone log for each phone call activity. Thus the mobile phone data having the phone numbers of individuals is a great resource for the unique identifiers of the users. The \textit{data-centric} social relational context can be generated from such unique identifiers, which represents one-to-one relationships according to individuals mobile phone data. Let's consider an example of an individual, her mother's phone number is well different with her friend's phone number and represents different relationships. Based on such uniqueness in individuals' mobile phone numbers in the dataset, the data-centric relationship values can be generated accordingly. Such generation always represents one-to-one relationships between individuals, even in same relationship category. For instance, in a friendship category, one's close friend's (one friend) phone number is well different with her best friend's (another friend) phone number, and generated different data-centric relationships based on their different numbers. Thus the main principle to determine the value of data-centric social relational context is ``each unique mobile phone number of an individual represents a particular one-to-one relationship with another individual''. 

Table \ref{sample-relationships}, shows a number of mobile phone numbers and corresponding data-centric social relational values. In Table \ref{sample-relationships}, we have shown the values of data-centric relationship (social context) as \{$Rel_{A}, Rel_{B},...,Rel_{E}$\}, according to the uniqueness in the given mobile phone numbers. For instance, mother's phone number (047XXXX231) represents one relation ($Rel_{A}$), while friend's phone number (047XXXX232) represents another relation ($Rel_{B}$) etc. Even this data-centric value is also able to distinguish the particular type of friendship. For instance, one friend's phone number (047XXXX232) represents one relation ($Rel_{B}$), while another friend's phone number (047XXXX234) represents another relation ($Rel_{D}$), which are shown in Table \ref{sample-relationships}. The corresponding sample semantic relationships are also shown in Table \ref{sample-relationships} for human understanding. The generated values of \textit{data-centric} social context shown in Table \ref{sample-relationships}, can be used in various context-aware mobile applications in order to assist the mobile phone users in their diverse phone call activities, according to their person-to-person relationships.

\section{Personalized Context-Aware Mobile Applications}
\label{Applications}
In this section, we discuss how the data-centric social relational context can be used in personalized mobile applications. The followings are the two real-life personalized applications for mobile phone users related to their phone call activities having the role of data-centric social relational context.

\subsection{Context-Aware Call Interruption Management}
Mobile phones are considered to be `always on, always connected' device but the mobile users are not always attentive and responsive to incoming communication \cite{chang2015investigating}. For this reason, sometimes people are often interrupted by incoming phone calls which not only create the disturbance for the phone users but also for the people nearby. Such kind of interruptions may create embarrassing situation not only in an official environment, e.g., meeting, lecture etc. but also affect in other activities like examining patients by a doctor or driving a vehicle etc. Sometimes these kind of interruptions may reduce worker performance, increased errors and stress in a working environment \cite{pejovic2014interruptme}. To circumvent this problem, people resort to switching their devices off altogether in environments such as meetings, or to configure silent mode that provide unobtrusive alerts of incoming calls which can be ignored if desired. However, these manual solutions have some shortcomings. First, missed calls are common, as the caller has no way of knowing before dialing whether the other party is available and willing to talk. Second, the recipient of the call typically has very limited information upon which to judge the importance of incoming calls. Thus an automated context-aware intelligent system is needed to automatically handle the incoming phone calls.

In the current state-of-the-art approaches, users need to define and maintain the social relationships and corresponding mobile phone configuring rules manually for their applications, which are static, i.e., the rules are not automatically discovered from the data. In general, users may not have the time, inclination, expertise or interest to maintain these rules manually in the real world \cite{sarker2016evidence}. In contrast, the data-centric social relational context can be discovered from individuals calling activity records in device logs, e.g., phone call logs, and corresponding mobile phone configuring rules can be generated using data mining techniques utilizing such mobile phone data. Using such rules based on the data-centric context can be used to build smart call interruption management systems, in order to adjust the modality of cell phone configuration, such as ringing to silent/vibrate or vice-versa. In such system, the mobile phone behaves differently from person-to-person according to the data-centric social relational context, which could be more effective to provide the personalized services intelligently.

\subsection{Context-Aware Call Reminder System}
Forgetting to make a phone call, i.e., outgoing call, is also a common problems in individual's daily life \cite{phithakkitnukoon2011behavior}. This could either be an event-based calls, such as making a phone call to plan for a meeting or seminar, or calling someone in a birth-day party or in a marriage anniversary, etc., or a nonevent-based calls, such as making a phone call to parents or significant persons at night or on weekends, calling girlfriend or boyfriend during a lunch break or dinner or a particular time periods in weekdays or weekends, calling someone in a particular location, e.g., calling only close friends in a particular playground or in a specific restaurant, etc. Thus a context-aware call reminder system is needed to intelligently remind the users to make a phone call to a particular person.

The data-centric social relational context can play a significant role to provide the personalized services in such context-aware system. The social relational context can be discovered from individuals calling activity records in device logs, e.g., phone call logs, and corresponding call reminder rules can be generated using data mining techniques utilizing such mobile phone data. Using such reminder rules based on the data-centric context can be used to build smart call reminder system, in order to intelligently searches the desirable contact from the large contact list and reminds a user to make a phone call to a particular person according to that context. Thus, the mobile phone behaves differently from person-to-person according to the data-centric social relational context, which could be more effective to provide the personalized services intelligently.

\section{Conclusion}
\label{Conclusion}
In this paper, we have explored the potentiality of data-centric social context, and briefly discussed how individuals' mobile phone data can be used as a social context in providing personalized services. We also discussed various categories of interpersonal relationships in our societies and their impacts and issues on individuals phone call activities. We further discussed how the data-centric social relational context or the interpersonal relationship between individuals can play a significant role in building context-aware personalized mobile applications, such as context-aware call interruption management system, context-aware smart call reminder system. We believe that the concept of data-centric social context could be more effective to provide various personalized services in human-centric context aware applications.

\bibliographystyle{plain}
\bibliography{bibfile/interruption}
\end{document}